\documentclass[12pt,a4paper]{article}
\usepackage[top=20truemm,bottom=20truemm,left=18truemm,right=18truemm]{geometry}

\usepackage{hyperref}
\usepackage{amsmath, amsthm}
\usepackage{amssymb}
\usepackage{multirow}
\usepackage{graphicx}

\newcommand{\Trp}{\mathrm{Tr}'}
\newcommand{\g}{\mathbf{g}}

\newcommand{\vo}{\mathrm{vol}_{\Sigma}}
\newcommand{\dvo}{d\mathrm{vol}_{\Sigma}}
\newcommand{\Acal}{\mathcal{A}}
\newcommand{\Sigmao}{\Sigma_{o}}
\newcommand{\Sigmab}{\overline{\Sigma}_{o}}
\newcommand{\Ncal}{\mathcal{N}}
\newcommand{\Psic}{\Psi_{c}}

\newcommand{\SU}{\mathrm{SU}}
\newcommand{\SO}{\mathrm{SO}}
\newcommand{\U}{{\mathrm{U}}}

\numberwithin{equation}{section}

\begin{document}
\begin{flushright}
OU-HET 846\\
IPMU14-0365
\end{flushright}
\begin{center}
{\LARGE Two-dimensional superconformal field theories \\ from Riemann surfaces with a boundary}\\
\vspace{5mm}
{\large Koichi Nagasaki${}^1$ and Satoshi Yamaguchi${}^2$}\\
\vspace{5mm}
{\small ${}^1$Kavli IPMU (WPI), University of Tokyo, 5-1-5 Kashiwanoha, Kashiwa, Chiba 277-8583, Japan \\
\texttt{kohichi.nagasaki@ipmu.jp}\\
${}^2$Department of Physics, Osaka University, 1-1 Machikaneyama, Toyonaka, Osaka 560-0043, Japan\\
\texttt{yamaguch@het.phys.sci.osaka-u.ac.jp}}\\
\end{center}
\begin{abstract}
We consider a 2-dimensional conformal field theory (CFT) obtained from twisted compactification of the 4-dimensional $\Ncal=4$ super Yang-Mills theory on a Riemann surface with boundary.
We find the boundary conditions for preserving some of the supersymmetry.
In particular an $\mathcal{N}=(2,2)$ superconformal field theory is obtained from supersymmetry breaking due to the boundary from $\mathcal{N}=(4,4)$.
In this case we calculate the central charge of the CFT and show its dependence on the topology of the Riemann surface.
\end{abstract}
%%%%%Sec 1.Introduction
\section{Introduction and summary}
We often find an interesting relationship between a geometry and a supersymmetric quantum field theory by compactifying a higher dimensional conformal field theory.  The class \textit{S} theories \cite{Gaiotto:2009we} are famous examples which are obtained by compactification of 6-dimensional (2,0) superconformal field theories (SCFTs) by Riemann surfaces.
Alday-Gaiotto-Tachikawa correspondence \cite{Alday:2009aq, Wyllard:2009hg} is a relation between  a class \textit{S} theory and a 2-dimensional CFT on the Riemann surface.
SCFTs obtained from a $d$-dimensional theory compactified on various manifolds are studied, for example,  \cite{Bershadsky:1995vm, Dimofte:2011ju, Bah:2011vv, Bah:2012dg, Klemm:1996bj, Cecotti:2011iy,Okazaki:2014sga}.

It is also interesting to consider a Riemann surface with boundary.
However for the class \textit{S} theories
%since they considered M5-branes wrapping on Riemann surfaces which cannot have a supersymmetric boundary,
it seems difficult to introduce a boundary of the Riemann surface since an M5-brane cannot have a supersymmetric boundary.
%If we consider an M5-brane, there does not such a brane that these M5-branes can end on.
%We would like to consider cases where a theory has a boundary.

In this paper we construct 2-dimensional CFTs obtained from compactification of 4-dimensional gauge theories on Riemann surfaces with boundary.
To realize a boundary theory we consider type IIB superstring in this paper.
Our gauge theory is a 4-dimensional $\Ncal=4$ super Yang-Mills theory (SYM) realized on the world-volume of D3-branes.
These D3-branes can end on D5-branes or NS5-branes, and thus can have a boundary.

The 2-dimensional CFTs obtained from compactification on closed Riemann surfaces \cite{Bershadsky:1995vm} are studied by using $c$-extremization \cite{Benini:2012cz, Benini:2013cda,Karndumri:2013iqa, Karndumri:2013dca}.
%When there are Abelian flavor symmetries, a mixing of R-symmetry $\U(1)_R$ and some of flavor symmetries can also become an R-symmetry. They proposed a method for finding the exact R-symmetry.
This method is 
%called the $c$-extremization and 
an analogue to $a$-maximization in 4-dimensions \cite{Intriligator:2003jj, Jafferis:2010un} and $F$-maximization in 3-dimensions \cite{Closset:2012vg}.
For $a$-maximization its gravity dual is studied in \cite{Martelli:2005tp, Martelli:2006yb, Butti:2005vn, Eager:2010yu, Tachikawa:2005tq}.

%The $c$-extremization is essential to find the correct result.
%Without it there is a mismatch between the supergravity and field theory results for the central charges \cite{Almuhairi:2011ws}.
%
%boundary

In this paper we study the 4-dimensional $\Ncal=4$ SYM on $\mathbb{R}^{1,1}\times \Sigmao$ where $\Sigmao$ is a Riemann surface with a boundary.  In the low energy limit this theory is expected to become a 2-dimensional CFT.  We find a class of boundary conditions at the boundary of $\Sigmao$ which preserve some of the supersymmetry, following the strategy of \cite{Gaiotto:2008sa}.  The boundary is a geodesic and preserves the $\Ncal=(0,1), \ (1,1),\ (2,2)$ supersymmetry out of the $\Ncal=(0,2), \ (2,2),\ (4,4)$ original bulk supersymmetry, respectively.  It is an interesting future work to study more general boundary conditions as in \cite{Gaiotto:2008sa,Gaiotto:2008sd,Gaiotto:2008ak,Hashimoto:2014vpa,Hashimoto:2014nwa} and S-duality.
In this paper we also show some attempt to find a different class of boundary conditions.
%which is called ``NS5-like'' condition.
%When one consider a theory with $\U(1)^n$ Abelian global symmetry group, there are Gauge and gravitational anomalies \cite{AlvarezGaume:1983ig}. 
%These anomalies give the 't Hooft anomaly coefficients.

Among these theories we calculate the central charge for the $\mathcal{N}=(2,2)$ case because in this case 
the central charge is related to the 't Hooft anomaly coefficients which are invariant under the renormalization group flow \cite{hooft1980recent}.  We obtain a positive central charge only when the Euler number $\chi_o$ of $\Sigmao$ is negative.  In this case the central charge is written as
\begin{align}\label{Eq:result}
c=3d_G |\chi_o|,
\end{align}
where $d_G$ is the dimension of the gauge group.  This theory has the $\Ncal=(2,2)$ superconformal symmetry 
with $c=3\times$(integer).  Therefore this theory seems to be a sigma model with a Calabi-Yau target space.
Further study of this theory, in particular the relationship with the theory of \cite{Bershadsky:1995vm}, is also an interesting problem.
This result coincides with the case for the central charge of theories compactified on the closed Riemann surfaces \cite{Benini:2012cz, Benini:2013cda, Maldacena:2000mw}.
Studying the reason of the coincidence between out result \eqref{Eq:result} and previous works \cite{Benini:2012cz, Benini:2013cda, Maldacena:2000mw} is an interesting future work.

Another interesting future work is to investigate the realization in the string theory and AdS/CFT correspondence \cite{Maldacena:1997re}.  Our setup is realized by D3-branes wrapping on a holomorphic cycle in a local Calabi-Yau manifold and ending on a 5-brane system
\cite{Hanany:1996ie,Kitao:1998mf, Gaiotto:2008sa,Gaiotto:2008sd,Gaiotto:2008ak,Hashimoto:2014vpa,Hashimoto:2014nwa}.

The construction of this paper is as follows:
In section 2 we introduce a twisted compactification of 4-dimensional gauge theories following \cite{2011arXiv1111.4234C, Bershadsky:1995qy, Maldacena:2000mw}.
In section 3 we find a condition for preserving supersymmetry and calculate the central charge.
%This condition imposes the vanishing of the normal component of the supercurrent at the boundary. 
%We find an example of the boundary condition which preserves an appropriate supersymmetry. 

%%%%%Sec 2.Twisted Compactification of $\mathcal{N}=4$ SYM
\section{Twisted compactification of $\mathcal{N}=4$ SYM}
We first review a 4-dimensional $\Ncal=4$ SYM on a curved spacetime following \cite{Benini:2012cz,Benini:2013cda}.
In subsection \ref{Subsec:Reduction}, first we obtain the action on the flat spacetime.
In subsections \ref{Subsec:RiemannSurf} and \ref{Subsec:Twist} we introduce a closed Riemann surface with constant curvature and twist the theory.
We also show how many supersymmetries are preserved by compactification on closed Riemann surfaces.

%%%Subsec 2.1
\subsection{$\mathcal{N}=4$ SYM on the flat spacetime}\label{Subsec:Reduction}
%We follow the notation of \cite{Benini:2013cda}.
%Let us begin with a 10-dimensional gauge theory with the following coordinates:
%\begin{equation}
%X^M,\:\: M=0,1,\cdots, 9.
%\end{equation}
%A 2-dimensional CFT lives on the flat space spanned by coordinates $X^0$ and $X^1$.
%$X^2$ and $X^3$ are coordinates on the Riemann surface and we redefine these coordinates as $(X^2, X^3) =: (x,y)$. We fix these coordinates such that the boundary is located at $y=0$.
The 4-dimensional $\mathcal{N}=4$ SYM action on the flat spacetime is obtained by the trivial dimensional reduction from the 10-dimensional SYM.
%First let us write a flat 4-dimensional action obtained by dimensional reduction.
It contains a 10-dimensional vector field $A_M,\ M=0,1,\dots,9$ and a 10-dimensional Majorana-Weyl spinor $\Psi$, which satisfies $\Gamma_{0123456789}\Psi=\Psi$. Both of them are in the adjoint representation of the gauge group $G$.  The vector field is decomposed into a 4-dimensional vector $A_{\mu},\ \mu=0,1,2,3, $ and 6 scalars $\Phi_{A}=A_{A},\ A=4,\dots,9$ in 4-dimensions.
The action is written as
\begin{equation}
S= \frac{1}{{g_{\rm YM}}^2} \int d^{4}x \Trp \left\{-
	\frac{1}{4}F_{MN}F^{MN} + \frac{i}{2}\overline{\Psi}\Gamma^M D_M \Psi \right\},
\end{equation}
where $g_{\rm YM}$ is the 4-dimensional gauge coupling.
$F_{MN}\ M,N =0,1,\dots ,9$ is defined as
\begin{align}
&F_{\mu\nu} = \partial_{\mu} A_{\nu} - \partial_{\nu} A_{\mu} + i[A_{\mu}, A_{\nu}],\qquad
\mu,\nu=0,1,2,3,\\
&F_{\mu A}=-F_{A\mu} = \partial_{\mu} \Phi_{A} + i[A_{\mu}, \Phi_{A}]=:D_{\mu} \Phi_{A},\\
&F_{AB}=i[\Phi_{A},\Phi_{B}].
\end{align}
The covariant derivative for $\Psi$ is defined as
\begin{align}
D_{\mu}\Psi=\partial_{\mu}\Psi+i[A_{\mu},\Psi],\qquad
D_{A}\Psi=i[\Phi_{A},\Psi].
\end{align}
$\Trp$ is  a trace normalized as $\Trp=\frac{1}{h^{\vee}} \mathrm{Tr}_{\text{adjoint}}$ where $h^{\vee}$ is the dual Coxeter number. For example, $\Trp=2\mathrm{Tr}_{\text{fundamental}}$ for $\SU (N)$.
The action is rewritten as
\begin{align}
S=\frac{1}{{g_{\rm YM}}^2}\int d^4x \Trp
\bigg\{-\frac{1}{4}F_{\mu\nu}F^{\mu\nu}
	- \frac12 D_\mu \Phi_A D^\mu \Phi^A
	+\frac{1}{4}[\Phi_A,\Phi_B][\Phi^A,\Phi^B] \nonumber\\
	+\frac{i}{2}\overline{\Psi}\Gamma^\mu D_\mu \Psi - \frac12\overline{\Psi}\Gamma^A [\Phi_A,\Psi]
\bigg \}.
\end{align}
This action is invariant under the supersymmetry transformation:
\begin{align}\label{SUSYTransf}
\delta A_M = i\bar{\epsilon}\Gamma_M \Psi,\:\:
\delta \Psi = \frac{1}{2}\Gamma^{MN}F_{MN}\epsilon.
\end{align}
The parameters $\epsilon$ are Majorana-Weyl fermions satisfying
\begin{equation}
\Gamma_{0123456789} \epsilon = \epsilon.
\end{equation}
Then the supersymmetry current is obtained as
\begin{equation}\label{SUSYCurrent}
J^{\mu}
= \frac{i}{2}\Trp \{2F^{\mu N}\Gamma_N-F_{KL}\Gamma^{KL \mu}\}\Psi
= \frac{i}{2}\Trp \{
F_{KL}\Gamma^{KL}\Gamma^{\mu} \Psi
\}.
\end{equation}

%To obtain the 4-dimensional theory we do dimensional reduction of the theory. The reduction to the 4-dimensional theory is defined by setting the partial derivatives for extra directions are  vanished: $\partial_A = 0$.
%On this resulted theory the external fields are interpreted as six scalar fields:
%\begin{equation}
%\Phi_A := A_A, \:\: A=4,5,\cdots, 9.
%\end{equation}
%Then, the action on the flat 4-dimensional space is
%where the covariant derivative for scalar fields and the fermion field is
%\begin{equation}\label{Eq:CovDerivPhiPsi}
%D_\mu \Phi_A = \partial_\mu \Phi_A + i[A_\mu, \Phi_A],\:\:
%D_\mu \Psi = \partial_\mu \Psi + i[A_\mu, \Psi].
%\end{equation}
%The Greek indices refer to 4-dimensional space-time where our gauge theory lives.

%%%Subsec 2.2
\subsection{Riemann surfaces}\label{Subsec:RiemannSurf}
%Riemann Surfaces are classified topologically into three types --- a sphere (${S}^2$), a torus (${T}^2$) or a hyperbolic surface (${H}^2$).
%In each case the metric on the Riemann surface is given as
%\begin{equation}
%ds^2=e^{2h(x,y)}(dx^2+dy^2),
%\end{equation}
%with
%\begin{align}
%h(x,y) := \begin{cases}
%-\ln\left(\frac{x^2+y^2+1}{2}\right) & (\mathbf{g}=0)  \\
%\frac{1}{2}\ln(2\pi) & (\mathbf{g}=1) \\
% -\ln(x) & (\mathbf{g}>1).
%  \end{cases}
%\end{align}

%Then the vielbeins are $E^2=e^h dx$ and $E^3=e^h dy$.
We will consider this 4-dimensional $\mathcal{N}=4$ SYM theory compactified on a compact Riemann surface $\Sigma$.  In this paper we concentrate on a Riemann surface with constant curvature $R=2\kappa$, where
\begin{align}
\kappa=\begin{cases}
\:\: +1\;\; &(\mathbf{g}=0)\\
\:\: 0\;\; &(\mathbf{g}=1)\\
\:\: -1\;\; &(\mathbf{g}>1),
\end{cases}
\end{align}
for a genus $\g$ closed Riemann surface.  We denote the coordinates of this Riemann surface by $(x^2,x^3)$, the vielbein by $E^a,\ a=2,3$, and the spin connection by $\Omega^{23}$.  The curvature 2-form is written as $R^{23}=d\Omega^{23}$, and thus the Gauss-Bonnet theorem reads
\begin{align}
\int_{\Sigma} d\Omega^{23}=\frac12\int_{\Sigma} \sqrt{g}R=4\pi(1-\g).
\end{align}
For $\g\ne 1$ the volume of the Riemann surface is
\begin{align}
\vo =4\pi|1-\g|,\label{volSigma}
\end{align}
and the volume form is
\begin{align}
\dvo=\kappa d\Omega^{23}.\label{Eq:SpinCurvature}
\end{align}

%%%Subsec 2.3
\subsection{Twisted gauge theory on the curved spacetime}\label{Subsec:Twist}
Now we consider the 4-dimensional $\mathcal{N}=4$ SYM theory on a curved spacetime with the metric $g_{\mu\nu}$ and a background $\SO (6)$ gauge field $\mathcal{A}_{\mu}=\frac12\mathcal{A}_{\mu}^{AB}M_{AB}$, where $M_{AB},\ A,B=4,\cdots,9$ are the $\SO (6)$ generators.  The action becomes
\begin{align}
S=\frac{1}{{g_{\rm YM}}^2}\int d^4x \sqrt{g}\Trp
\bigg\{-\frac{1}{4}F_{\mu\nu}F^{\mu\nu}
	- \frac12 D'_\mu \Phi_A D'^\mu \Phi^A
	+\frac{1}{4}[\Phi_A,\Phi_B][\Phi^A,\Phi^B]\nonumber\\
	+\frac{i}{2}\overline{\Psi}\Gamma^\mu D'_\mu \Psi - \frac12\overline{\Psi}\Gamma^A [\Phi_A,\Psi]
\bigg \},
\end{align}
where the covariant derivative $D'_{\mu}$ includes the spin connection and the $\SO (6)$ gauge field
\begin{align}
D'_\mu \Phi_A &:= \partial_\mu \Phi_A + i[A_\mu, \Phi_A] + \sum_B \mathcal{A}_{\mu}^{AB}\Phi_B,\label{Eq:TwDerivPhi}\\
D'_\mu \Psi &:= \partial_\mu \Psi +i[A_{\mu},\Psi]+ \frac{1}{4}\Omega_{\mu}^{ab}\Gamma_{ab} -i \mathcal{A}_\mu \Psi.
\end{align}
Here $\mathcal{A}_{\mu}\Psi:=\frac{i}{4}\mathcal{A}_{\mu}^{AB}\Gamma_{AB}\Psi$.
In order to preserve the supersymmetry, a parameter of the supersymmetry transformation \eqref{SUSYTransf} should satisfy the Killing spinor equation.
The twisted Killing spinor equation is
\begin{equation}\label{Eq:KillingSpEq}
 D'_\mu \epsilon :=
  \left(
  \partial_\mu + \frac{1}{4}\Omega_{\mu}^{ab}\Gamma_{ab} -i \mathcal{A}_\mu
  \right) \epsilon = 0.
\end{equation}
%The relation of the spin connection, $\Omega_{23} := \Omega_{23,\mu} dx^{\mu}$, and the curvature is
%\begin{equation}\label{Eq:SpinCurvature}
%\int_{\Sigma_{\bf g}} d\Omega_{23}
%= \frac{1}{2}\int_{\Sigma_{\bf g}} R\:d{\rm vol_{\Sigma_{\bf g}}}
%= \kappa \int_{\Sigma_{\bf g}} d{\rm vol_{\Sigma_{\rm g}}},\hspace{1cm}
%\therefore d\Omega_{23}=\kappa d{\rm vol_{\Sigma_{\rm g}}}.
%\end{equation}

We choose the external gauge field $\mathcal{A}_\mu$ in $\SO (2)^3 \subset \SO (6)$, such that the field strength,
\begin{equation}
F=d\mathcal{A}, \:\: \mathcal{A}=\mathcal{A}_\mu dx^\mu,
\end{equation}
satisfies
\begin{align}\label{Eq:fieldstrength}
F=\begin{cases}
\:\: -T \:d{\rm vol_{\Sigma}}\;\; &(\mathbf{g}\neq 1)\\
\:\: -T\frac{2\pi}{\rm vol_{\Sigma}} d{\rm vol_{\Sigma}}\;\; &(\mathbf{g}=1).
\end{cases}
\end{align}
Here $T$ is an $\SO(2)^3$ generator
\begin{equation}\label{Eq:TwistSO3}
T=a_1T_1+a_2T_2+a_3T_3,
\end{equation}
where $a_i$ are parameters of twisting and $T_i,\ i=1,2,3$ are generators expressed in the spinor representation
\begin{align}
   T_1=\frac{i}{2}\Gamma^{45},\:\:
   T_2=\frac{i}{2}\Gamma^{67},\:\:
   T_3=\frac{i}{2}\Gamma^{89}.
\end{align}
%By these generators the external field $\mathcal{A}_\mu$ is decomposed as
%\begin{align}
%\mathcal{A}
%= \mathcal{A}_\mu dx^{\mu}
%= \sum_{A,B} \frac{1}{2} \mathcal{A}_{\mu, AB} \left(\frac{i}{2}\Gamma^{AB}\right).
%\end{align}

The condition for existing covariantly constant spinors is, from eq.~\eqref{Eq:KillingSpEq},
\begin{align}
D'_{\mu}\epsilon =0 \:\Rightarrow\:
[D'_2,D'_3]\epsilon =0 \:\Rightarrow\:
\left(
\frac{1}{2}d\Omega_{23}\Gamma^{23} -i d\mathcal{A}\right)\epsilon =0.
\end{align}
Using the relations \eqref{Eq:SpinCurvature} and \eqref{Eq:fieldstrength},
\begin{equation}
\left(
\frac{1}{2}\kappa\: d\vo\cdot\Gamma^{23} +i d\vo\cdot T
\right)\epsilon =0.
\end{equation}
Finally, substituting eq.\eqref{Eq:TwistSO3}, the supersymmetry condition is
\begin{equation}
\left(
-{\kappa}i\Gamma^{23}+a_1i\Gamma^{45}+a_2i\Gamma^{67}+a_3i\Gamma^{89}
\right)\epsilon =0.
\label{bulkEpsilonCondition}
\end{equation}
The amount of the supersymmetry depends on the number of the non-zero parameters among $a_i,$ $i=1,2,3$.
Let us classify them here:

\begin{enumerate}
\item{All $a_i$ are non-zero:}
$\left(
-{\kappa}\Gamma^{23}+a_1\Gamma^{45}+a_2\Gamma^{67}+a_3\Gamma^{89}
\right)\epsilon =0$.\\
In this case the number of the supersymmetries is $\mathcal{N}=(0,2)$. The constraint for the parameters $a_i$ is
\begin{equation}
a_1+a_2+a_3 =\kappa.
\end{equation}

\item{Two of $a_i$ are non-zero:}
$\left(
-\kappa\Gamma^{23}+a_1\Gamma^{45}+a_2\Gamma^{67}
\right)\epsilon =0$.\\
In this case the number of the supersymmetries is $\mathcal{N}=(2,2)$. The constraint for the parameters $a_i$ is
\begin{equation}
a_1+a_2 =\kappa.
\end{equation}

\item{One of $a_i$ is non-zero:}
$\left(
-\kappa\Gamma^{23}+a_1\Gamma^{45}
\right)\epsilon =0$.\\
In this case the number of the supersymmetries is $\mathcal{N}=(4,4)$. The constraint for the non-zero parameter $a_1$ is
\begin{equation}
a_1 = \kappa.
\end{equation}

\item{No background field:}
$\left(
-{\kappa}\Gamma^{23}
\right)\epsilon =0$.\\
In this case the number of the supersymmetries is $\mathcal{N}=(8,8)$. This situation is realized only for the zero curvature case $\kappa=0$, i.e. $\mathbf{g}=1$.
\end{enumerate}
These results are summarized in Table \ref{Table:GenusSUSY}.

\begin{table}[h]
\centering
  \begin{tabular}{| c | c | c |}
    \hline
    \# of $a_i\neq 0$ & $\Ncal$ & \;\;\;\;\;\;\;\; $\mathbf{g}$ \;\;\;\;\;\;\;\;  \\ \hline
    3 & $(0,2)$ & all  \\ \hline
    2 & $(2,2)$ & all  \\ \hline
    1 & $(4,4)$ & all  \\ \hline
    0 & $(8,8)$ & $1$\\ \hline
  \end{tabular}
\caption{Remaining supersymmetries for closed Riemann surfaces.}\label{Table:GenusSUSY}
\end{table}

%%%%%Sec 3.SUSY Boundary Condition and Central Charge
\section{Supersymmetric boundary condition and central charge}
In this section we introduce a boundary on the Riemann surface.
We assume that the boundary is a geodesic.
First we explain this assumption is appropriate and simplifies our argument.
After that we study the boundary condition for preserving some supersymmetries.
We obtain the central charge when the $\Ncal=(2,2)$ supersymmetry is preserved.
In this calculation we assume that the two-dimensional theory at low energies is conformal. 
However, if the calculation gives the negative central charge then this indicates the assumption is violated. 
We also show an attempt to find other class of boundary conditions.

\subsection{Shape of the Boundary}
In this paper we focus on Riemann surfaces with one boundary.
We also assume that these surfaces have constant curvature.
%These Riemann surface and its metric is realized as shown in Subsection \ref{Subsec:RiemannSurf}.
In this paper we only consider a geodesic boundary for simplicity.
There could be a non-geodesic boundary which preserves some supersymmetry, although we do not find an example.
The analysis is rather simple for the geodesic boundary for the following reasons.
Let $(x^2, x^3)$ the coordinates of the Riemann surface and the geodesic boundary $x^3=0$.
Then we can choose a gauge such that locally $\Acal_{2}=\Acal_{3}=0$ on the boundary
since $\Acal$ is proportional to $\Omega^{23}$ and we can choose the gauge $\Omega^{23}=0$ on a geodesic.
Then terms including the external gauge field $\mathcal{A}_\mu$ in the covariant derivative \eqref{Eq:TwDerivPhi} can be omitted and $D'=D$ is satisfied at least locally.
However we cannot ignore the holonomy along the boundary.  The boundary condition must be consistent with this holonomy.
Another reason for choosing the geodesic boundary is that we want to use the doubling trick later.
If the boundary is a geodesic, one can join together the Riemann surface and a copy of it with the opposite orientation to construct a closed surface with constant curvature.

%The external gauge field $\mathcal{A}$  can be locally set to zero.
%But globally this gauge field has a physical meaning.
Let us see the holonomy of this external gauge field along the boundary.
First for simplicity we consider an $S^2$ with a boundary at the equator --- a northern (or southern) hemisphere $S^2_{+}$ ($S^2_{-}$). This holonomy is given by
\begin{equation}\label{Eq:IntGaugefield}
\oint_{\partial S^2_{+}} \mathcal{A}
= \int_{S^2_{+}} d\mathcal{A}
= \int_{S^2_{+}} F = \text{Magnetic flux}.
\end{equation}
Here we use Stokes' theorem to express it as an integral of the gauge field strength. This integral gives a magnetic flux through the surface $S^2_{+}$. Due to the Dirac quantization condition, the integral of magnetic flux on the $S^2$ is an integral multiplication of $2\pi$.
Now this gauge field is distributed isotropically. Then the integral only over the northern hemisphere \eqref{Eq:IntGaugefield} gives an integer or a half integer times $2\pi$.
We can use the same strategy for a general Riemann surface $\Sigmao$ with one geodesic boundary.
Let $\Sigma$ be the closed Riemann surface made by gluing $\Sigmao$ and a copy of it with the opposite orientation $\Sigmab$ along their boundaries.  Notice
that the genus $\g$ of $\Sigma$ is an even number and thus it is not $1$.
The holonomy along this boundary can be written by using eqs.~\eqref{Eq:fieldstrength}, \eqref{Eq:TwistSO3}, \eqref{volSigma} as
\begin{equation}
H:=\exp \left( i\oint_{\partial\Sigmao} \mathcal{A} \right)
=\exp \left( i\int_{\Sigmao} F \right)
=\exp \left( \frac{i}{2}\int_{\Sigma} F \right)
= \prod_{i=1,2,3}\exp (-i\pi n_i T_i),
\label{holonomy}
\end{equation}
where $n_i:=2|1-\g|a_i$ are integers \cite{Benini:2013cda}.  Later we use the fact 
\begin{align}
H^2=\exp \left( i\int_{\Sigma} F \right)=1 \label{H21}
\end{align}
following from the Dirac quantization condition.
%Then the directions can be rotated only $\pi$ when we take it around the boundary line.
%We conclude that $\mathcal{A}=0$ is globally consistent.
The boundary condition considered in this paper later \eqref{Eq:XNeumanYDirichlet} is consistent with this holonomy \eqref{holonomy}.

\subsection{Boundary condition}
%We introduce a boundary to the Riemann surface.  As discussed in the previous subsection, we restrict this boundary to be the geodesic.
Let us here consider the boundary conditions which preserve some part of the supersymmetry.
%We expect this introduction breaks the supersymmetry because the introduction of the boundary breaks the translation symmetry normal to the boundary. Then this theory does not preserve all of the original supersymmetry.
%We use almost the same notation as in \cite{Gaiotto:2008sa,Hashimoto:2014vpa}.
For preserving the supersymmetry the current component normal to the boundary must be zero at the boundary ($x^3=0$). From eq.~\eqref{SUSYCurrent} this condition is expressed as
\begin{equation}\label{Eq:SupercurrentCond}
\overline{\epsilon}J^3=0
\:\:\Leftrightarrow\:\:\:
\Trp \left(\overline{\epsilon}F_{KL}\Gamma^{KL}\Gamma^3\Psi\right)=0.
\end{equation}
In this condition we can replace $D'$ by $D$ since we can choose the gauge where $\Acal=0$ at the boundary.
Thus we can employ the same strategy as \cite{Gaiotto:2008sa} (see also \cite{Hashimoto:2014vpa}).
Define the following matrices:
\begin{align}
&B_0 = \Gamma^{468579},\\
&B_1 = \Gamma^{3468},\\
&B_2 = \Gamma^{3579},
\end{align}
and redefine the scalar fields
\begin{align}
(X_4, X_6, X_8) &:= (\Phi_4, \Phi_6, \Phi_8),\\
(Y_5, Y_7, Y_9) &:= (\Phi_5, \Phi_7, \Phi_9).
\end{align}

%In the following three subsections (\ref{Subsec:02case}, \ref{Subsec:22case} and \ref{Subsec:44case}), we examine some cases where the original supersymmetries are $\mathcal{N} = (0,2)$, $(2,2)$ and $(4,4)$.
%Especially in $(4,4)$ case the exact R-symmetry is automatically
%determined and the central charge is related to the 't Hooft anomaly coefficients.

The boundary condition \eqref{Eq:SupercurrentCond} is decomposed into the following equations as done in \cite{Gaiotto:2008sa}:
%by the dependence of the Lorentz transformation:
\begin{align}
\Trp \overline{\epsilon}(\Gamma^{\mu\nu}F_{\mu\nu}+2\Gamma^{3\mu}F_{3\mu})\Gamma^3\Psi &= 0,\nonumber\\
\Trp \overline{\epsilon}(2\Gamma^{3a}D_3X_a+\Gamma^{ab}[X_a,X_b])\Gamma^3\Psi &= 0,\nonumber\\
\Trp \overline{\epsilon}(2\Gamma^{3m}D_3Y_m+\Gamma^{mn}[Y_m,Y_n])\Gamma^3\Psi &= 0,\nonumber\\
\Trp \overline{\epsilon}\Gamma^{\mu a}D_\mu X_a \Gamma^3\Psi &= 0,\nonumber\\
\Trp \overline{\epsilon}\Gamma^{\mu m}D_\mu Y_m \Gamma^3\Psi &= 0,\nonumber\\
\Trp \overline{\epsilon}\Gamma^{am}[X_a,Y_m] \Gamma^3 \Psi&= 0,
\label{currentConditionDecomposed}
\end{align}
where $\mu,\nu=0,1,2,$ $a,b=4,6,8,$ and $m,n=5,7,9$.
An example of the boundary condition is the NS5-brane like boundary condition
\begin{align}\label{Eq:XNeumanYDirichlet}
D_3 X_a = 0,\: Y_m = 0,\: F_{\mu 3} = 0,
\end{align}
for the bosonic fields.
%\begin{equation}
%F_{\mu 3}=0,
%\end{equation}
%or the Dirichlet boundary condition for the D5-brane
%\begin{equation}
%F_{\mu\nu}=0 \:\: (\mu\neq 3) \leftrightarrow \epsilon^{\mu\nu\lambda}F_{\nu\lambda}=0.
%\end{equation}
%
%For the fermion field we introduce the decomposition $\Gamma_3 \Psi =  \Psi' \otimes \vartheta$, where $\vartheta$ is an eigenvector of $B_2$:
For the fermionic fields we impose
\begin{equation}
B_2 {\Gamma^3 \Psi} = - \Gamma^3 \Psi
\label{NS5fermion}
\end{equation}
at the boundary.  Actually the NS5-brane like boundary conditions \eqref{Eq:XNeumanYDirichlet} and \eqref{NS5fermion} preserve the supersymmetry if the parameter $\epsilon$ satisfies
\begin{align}
B_2\epsilon = \epsilon.
\label{boundarySUSYepsilon}
\end{align}
The conditions \eqref{currentConditionDecomposed} is verified.  The condition \eqref{boundarySUSYepsilon} for $\epsilon$ kills half of the supersymmetry as follows.  If an $i\Gamma^{23}$ eigenvector $\epsilon_1$ satisfies \eqref{bulkEpsilonCondition}, $B_2 \epsilon_1$ also satisfies \eqref{bulkEpsilonCondition} and they are independent. Therefore among the linear combinations of these two independent parameters, one combination
$\epsilon=(1+B_2)\epsilon_1$ satisfies the condition \eqref{boundarySUSYepsilon}.  Since $\epsilon_1$ and $B_2\epsilon_1$ have the same chirality ($\Gamma^{01}$ eigenvalue), the preserved supersymmetry is as follows.
\begin{enumerate}
\item $\Ncal=(0,2)$ bulk $\Rightarrow$ $\Ncal=(0,1)$.
\item $\Ncal=(2,2)$ bulk $\Rightarrow$ $\Ncal=(1,1)$.
\item $\Ncal=(4,4)$ bulk $\Rightarrow$ $\Ncal=(2,2)$.
\end{enumerate}

Let us verify the boundary conditions \eqref{Eq:XNeumanYDirichlet} and \eqref{NS5fermion} are consistent with the holonomy \eqref{holonomy}.  For the vector representation $(H\Phi)_A=\pm \Phi_A$, so the conditions for the bosons \eqref{Eq:XNeumanYDirichlet} are consistent.  The consistency of the condition for the fermions \eqref{boundarySUSYepsilon} is verified by \eqref{H21} $H^2=1$ and $B_2 \Gamma^3 H \Psi=H^{-1} B_2 \Gamma^3 \Psi$.
%$H^2$ is written as
%\begin{align}
%H^2=\exp\left(-2\pi i (n_1 T_1 + n_2 T_2 +n_3 T_3)\right).
%\end{align}
%Notice that the eigenvalues of $T_i$ are $\pm \frac12$ in the spinor representation.
%We also have the relation $n_1+n_2+n_3 = 2\kappa |\g-1|$ and thus all $\frac12 (\pm n_1\pm n_2 \pm n_3)$ are integers. Therefore we conclude that $H^2=1$.

\subsection{$\mathcal{N}=(4,4)$ case and the central charge}\label{Subsec:central-charge}
The case where the bulk $\Ncal=(4,4)$ supersymmetry is broken to $\mathcal{N}=(2,2)$ by the boundary 
is interesting because of the R-symmetry of the $\Ncal=2$ superconformal symmetry.  In this case $a_2=a_3=0$ and $a_1=\kappa$, and $T$ in eq.~\eqref{Eq:TwistSO3} becomes
\begin{equation}
T=\kappa \frac{i}{2}\Gamma^{45}. \label{N44T}
\end{equation}
The preserved supersymmetry parameters satisfy eq.~\eqref{bulkEpsilonCondition}, which is rewritten as
\begin{align}
\Gamma^{2345}\epsilon=-\epsilon.
\end{align}
Then the exact central charge is obtained from the 't Hooft anomaly coefficient as in \cite{Benini:2013cda}.  However in our case the situation is much simpler since there is only one candidate $\U (1)$ symmetry $Q^R$ for the R-symmetry
\begin{equation}
Q^R = \frac{i}{2}\Gamma^{68} + \frac{i}{2}\Gamma^{79}.
\end{equation}
This is determined such that for the right moving supersymmetry parameters $\epsilon$  ($\Gamma^{01}\epsilon=+\epsilon$) satisfy $Q^{R}\epsilon=\pm \epsilon$ and the left moving ones satisfy $Q^{R}\epsilon=0$.   The right moving central charge is expressed as
%\begin{equation}
%\nabla^\mu J^I_\mu = \sum_L \frac{k^{IL}}{8\pi}F^{IL}_{\mu\nu}\epsilon^{\mu\nu}, \qquad
%c_R = 3k^{RR},
%\end{equation}
%where $I, L =1,2,\cdots, n$ and $n$ is the number of Abelian flavor symmetry groups.
%$k^{RR}$ is calculated by the R-current as
\begin{equation}
c = 3 \text{Tr}_\text{Weyl fermion}(\Gamma^{01}(Q^R)^2).
\end{equation}
In the above expression ${\rm Tr}_{\text{Weyl fermion}}$ means counting the number of the 2-dimensional Weyl fermions.
\begin{figure}[b]
    \centering
    \includegraphics[width=8cm]{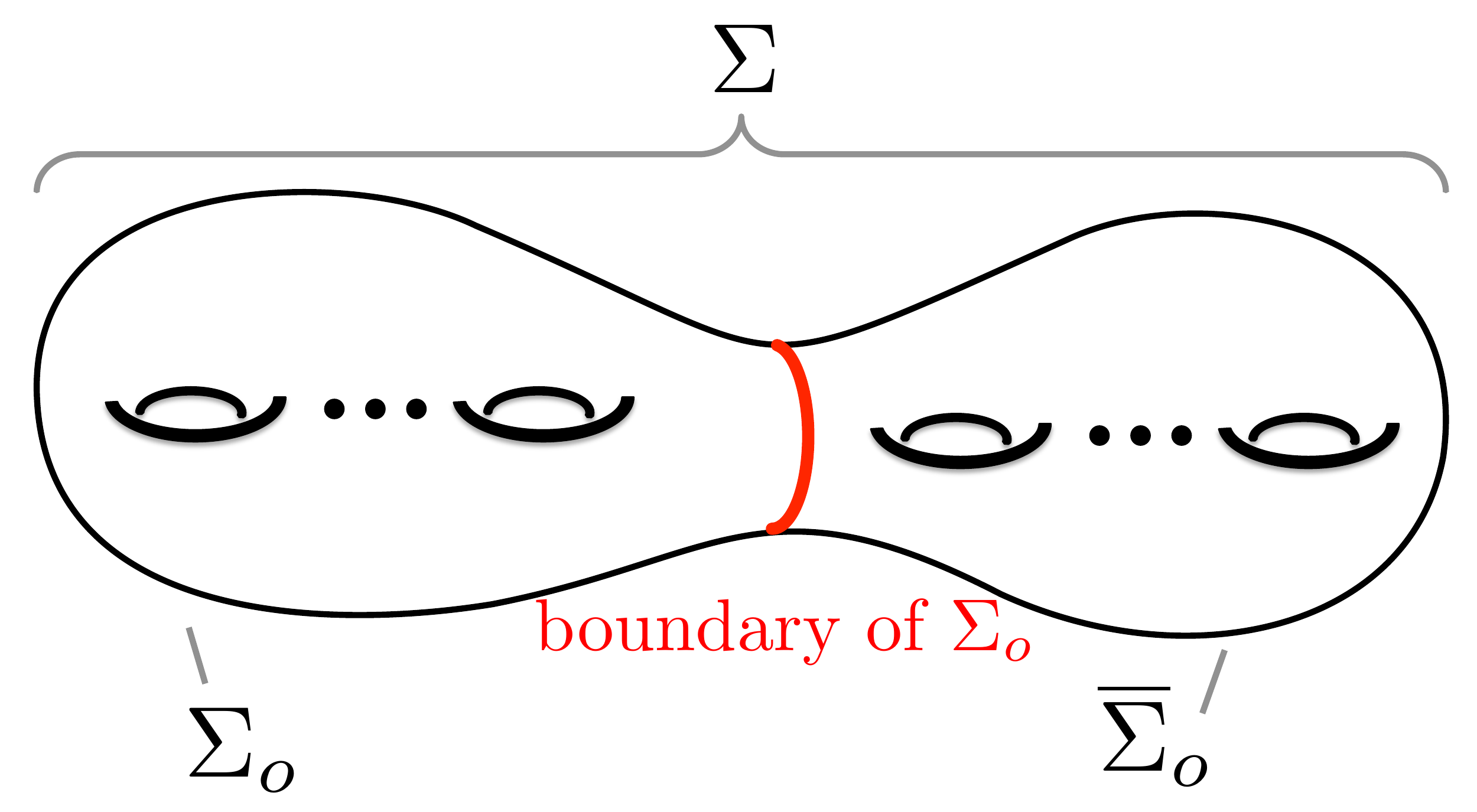}
    \caption{(Doubling trick) We construct a closed surface $\Sigma$ by taking $\Sigma_{o}$ and one with the opposite orientation $\Sigmab$.}
    \label{fig:DoubledSurface}
\end{figure}

%On the above situation we would like to calculate the central charge.
%%%doubling trick
The number of the chiral fermions can be counted by the index theorem as in \cite{Benini:2012cz}.  In this paper we use the doubling trick to map the problem to the index theorem in the closed Riemann surface.  We take the Riemann surface $\Sigmao$ and a copy with the opposite orientation $\Sigmab$, and join them together, $\Sigmao \bigcup \Sigmab=:\Sigma$, so that their boundaries are the same (See Figure \ref{fig:DoubledSurface}).
%The fermion fields must agree on the boundary:
%\begin{equation}\label{Eq:AgreeFermion}
%B_2\Psi = \Psi.
%\end{equation}
Originally $\Psi$ includes four 4-dimensional Weyl spinors.
Half of them satisfying $\Gamma^{6789}\Psi=-\Psi$ have charge $\pm 1$ of $Q^R$ and the others are neutral.
Let us denote these two charged 4-dimensional Weyl spinors $\Psi_{\pm}$ which satisfy $i\Gamma^{45}\Psi_{\pm}=\pm\Psi_{\pm}$ and $B_2\Psi_{\pm}=\Psi_{\mp}$.
%By taking the eigenstates summarized in Table \ref{Table:44EV} as a basis, $|\lambda^0,\cdots,\lambda^8\rangle$, $\Psi$ is decomposed as
%\begin{equation}\label{Eq:ExpansionPsi}
%\Psi = \Psi_{1} (z) |+----\rangle + \Psi_{2} (z) |+++--\rangle +\cdots.
%\end{equation}
%The matrix $B_2$ acts as these basis as follows:
%\begin{align}
%B_2 | + - - - - \rangle &= | + + + - - \rangle, \nonumber\\
%B_2 | - + - - - \rangle &= | - - + - - \rangle, \nonumber\\
%B_2 | + - - +  + \rangle &= | + + + + + \rangle, \nonumber\\
%B_2 | - + - +  + \rangle &= | - - + + + \rangle.
%\end{align}
These two fermions on $\mathbb{R}^{1,1}\times \Sigmao$ are treated as a fermion $\Psic$ on
$\mathbb{R}^{1,1}\times \Sigma$.  $\Psic$ is defined as
%Then the agreement condition \eqref{Eq:AgreeFermion} is equivalent to defining a field
\begin{align}
\Psic =
\begin{cases}
& \Psi_{-}(z),\quad (\text{Im}(z)\geq 0)\\
& \Psi_{+} (z^*), \quad (\text{Im}(z)\leq 0).
\end{cases}
\label{defPsic}
\end{align}
Here we use the complex coordinate $z=x^2+ix^3$ of $\Sigma$ such that
$\Sigmao$ is parametrized by Im$z\ge 0$,
$\Sigmab$ is parametrized by Im$z\le 0$ and $z\to z^{*}$ is the symmetry which exchanges $\Sigmao$ and $\Sigmab$.
Actually this $\Psic$ is continuous at the boundary due to the boundary condition \eqref{NS5fermion}.
Furthermore, we define extended spin connections and gauge fields
\begin{align}
\Omega^{23}_{\bar{z}} (z) :=
\begin{cases}
\Omega^{23}_{\bar{z}}(z) & (\text{Im}(z)\geq 0)\\
-\Omega^{23}_{z} (z^*) & (\text{Im}(z)\leq 0),
\end{cases}\qquad
\mathcal{A}^{45}_{\bar{z}} (z) :=
\begin{cases}
\mathcal{A}^{45}_{\bar{z}}(z) & (\text{Im}(z)\geq 0)\\
-\mathcal{A}^{45}_{z} (z^*) & (\text{Im}(z)\leq 0).
\end{cases}
\end{align}
%%%Dirac eq
Then according to the above definitions, the Dirac equations for $\Psi_{\pm}$ on $\mathbb{R}^{1,1}\times\Sigmao$ are equivalent to the one for $\Psic$ on $\mathbb{R}^{1,1}\times\Sigma$
\begin{align}
\Gamma^{\mu}D'_{\mu}\Psic (z) = 0.
\end{align}

We denote the number of 2-dimensional right(left)-moving massless fermions by $n_{R(L)}$ for the 4-dimensional
Weyl fermion $\Psic$.
%The Majorana-Weyl fermion field $\Psi$ is an 8 components spinor.
%We take a linear combinations
%\begin{align}
%&b_0 = \frac{1}{2}(-\Gamma^0 + \Gamma^1),\:
%b_0^{\dagger} = \frac{1}{2}(\Gamma^0 + \Gamma^1)\\
%&b_i = \frac{1}{2}(\Gamma^{2i} + i\Gamma^{2i+1}),\:
%b_i^{\dagger} = \frac{1}{2}(\Gamma^{2i} - i \Gamma^{2i+1})
%\end{align}
%Now we take the following basis for which $\Psi$ is expanded as \eqref{Eq:ExpansionPsi}:
%\begin{align}
%|\pm,\pm,\pm,\pm,\pm \rangle,\:\:
%\text{with } b_a |-,-,-,-,-\rangle =0, a=1,2,3,4.
%\end{align}
%The number of right and left moving fermions labelled by $\sigma$ is expressed as
%\begin{equation}
%n_R^{\sigma} = \# (|+,-,-,-,-\rangle),\:\:
%n_L^{\sigma} =  \# (|-,+,-,-,-\rangle).
%\end{equation}
%%%Index theorem
%The index theorem gives the difference of these numbers:
The index theorem 
%\cite{APSindexTh} 
gives the difference of these numbers and it is rewritten using eqs.~\eqref{Eq:fieldstrength}, \eqref{volSigma}
:
\begin{equation}
n_{R} - n_{L}
= -\frac{1}{2\pi}\int_{\Sigma} \mathrm{Tr}_{\Psic} F
= t 2|\g-1|,
\end{equation}
where $\mathrm{Tr}_{\Psic}$ is taken in the representation of $\Psic$, and
%we integrate over the closed surface ($\Sigma=\Sigmao\bigcup \Sigmao$) and
$t$ is the eigenvalue of $T$ for the fermion $\Psic$ which is given by $t = -\kappa/2$ using eqs.~\eqref{defPsic}, \eqref{N44T}.
%In this expression $\mathbf{g}$ means the genus number of the original Riemann surface which has a boundary.
%This number becomes $2\mathbf{g}$ for the closed surface constructed by the doubling trick.
%%%Result
Taking the multiplicity of the Lie algebra into account we obtain the result
\begin{align}\label{Eq:CentralCharge}
c &= 3d_G (n_R - n_L)\nonumber\\
&= -3d_G \kappa |\mathbf{g}-1|,
\end{align}
%In the above expression we used the fact $\kappa=-1$, for  $\g>1,$ after the doubling trick.
where $d_G$ is the dimension of the gauge group.
This expression gives the positive $c$ only when $\kappa=-1,\ (\g>1)$. In this case
\begin{align}
c= 3d_G |\chi_o|.
\end{align}
In the final expression we use the Euler number of the original Riemann surface with the boundary.
We are now considering a case where the Riemann surface has only one boundary, $\mathbf{b}=1$.
Then the Euler number of the original surface, $\Sigmao$, is $\chi_o = 2-2\mathbf{g}/2-\mathbf{b} = 1-\mathbf{g}$.

\subsection{Candidate for other types of boundary condition}
In this subsection, 
%In the following three subsections (\ref{Subsec:02case}, \ref{Subsec:22case} and \ref{Subsec:44case}), 
we examine boundary conditions different from the NS5-like shown in the previous subsections.
% \ref{Sec:BoundaryCond}.
We show some cases where the original bulk supersymmetries are $\mathcal{N} = (0,2)$, $(2,2)$ and $(4,4)$.
We study how these supersymmetries are broken when introducing the boundary.
%We show how many supersymmetries survive and its conditions.

In this subsection we use the following notation for the supersymmetry parameters $\epsilon_{I},\ I=1,\cdots,8$.
We diagonalize $\Gamma^{01}, i\Gamma^{M,M+1},\ M=2,4,6,8$ and denote the eigenvalues as follows:
\begin{align}
  \left\{
    \begin{array}{l}
      \Gamma^{01} \epsilon_I = \lambda^0_I \epsilon_I \\
      i\Gamma^{M,M+1} \epsilon_I = \lambda^M_I \epsilon_I
    \end{array}
  \right.\qquad\Rightarrow\qquad
  \left\{
    \begin{array}{l}
      \overline{\epsilon_I}\Gamma^{01} = -\lambda^0_I\overline{\epsilon_I} \\
      \overline{\epsilon_I} (i\Gamma^{M,M+1}) = -\lambda^M_I \overline{\epsilon_I}
    \end{array}
  \right. ,
\end{align}
where eigenvalues $\lambda^0_I, \lambda^M_I$ take values $+1$ or $-1$ and are summarized in Table \ref{44EV}.

\begin{table}[h]
\centering
  \begin{tabular}{| l  c || c | c | c | c | c |}
    \hline
     & & $\lambda^{0}_{I}$ & $\lambda^{2}_{I}$ & $\lambda^{4}_{I}$ & $\lambda^{6}_{I}$ & $\lambda^{8}_{I}$ \\ \hline
   \multirow{8}{*}{$I$} & 1 & $-$ & $+$ & $+$ & $+$ & $+$ \\
   				& 2 & $-$ & $-$ & $-$ & $-$ & $-$  \\
      				& 3 & $+$ & $+$ & $+$ & $+$ & $-$  \\
      				& 4 & $+$ & $-$ & $-$ & $-$ & $+$  \\ \cline{3-7}
   				& 5 & $-$ & $+$ & $+$ & $-$ & $-$ \\
      				& 6 & $-$ & $-$ & $-$ & $+$ & $+$  \\
   				& 7 & $+$ & $+$ & $+$ & $-$ & $+$  \\
      				& 8 & $+$ & $-$ & $-$ & $+$ & $-$  \\ \hline
  \end{tabular}
\caption{Eigenvalues of $\epsilon_I$.}\label{44EV}
\end{table}

\subsubsection{$\mathcal{N}=(0,2)$ case}\label{Subsec:02case}
The supersymmetry parameters preserved in the bulk are $\epsilon_1$ and $\epsilon_2$.

The current condition \eqref{Eq:SupercurrentCond} for these generators is
\begin{align}
& \Trp \:\overline\epsilon\left(
F_{01}\Gamma^{01}+F_{23}\Gamma^{23}+F_{45}\Gamma^{45}+F_{67}\Gamma^{67}+F_{89}\Gamma^{89}
\right)\Gamma^3\Psi = 0\label{02SUSYCond1},\\
& \Trp \:\overline\epsilon\left(
F_{M,N}\Gamma^{M,N}+F_{M,N+1}\Gamma^{M,N+1}+F_{M+1,N}\Gamma^{M+1,N}+F_{M+1, N+1}\Gamma^{M+1, N+1}
\right)\Gamma^3\Psi = 0\label{02SUSYCond2},\\
&\hspace{3cm} (M,N) = (0,2), (0,4), (0,6), (0,8), (2,4), (2,6), (2,8), (4,6), (4,8), (6,8).\nonumber
\end{align}
We impose the boundary condition for the fermion field:
\begin{equation}
-i\Gamma^{23}\Psi=\Psi.
\end{equation}

From the first equation \eqref{02SUSYCond1},
\begin{align}
&\Trp \:\overline{\epsilon_I}\left(
F_{01}\Gamma^{01}+F_{23}\Gamma^{23}+F_{45}\Gamma^{45}+F_{67}\Gamma^{67}+F_{89}\Gamma^{89}
\right)\Gamma^3(-i\Gamma^{23}\Psi) = 0\nonumber\\
&\leftrightarrow
\Trp \:\overline{\epsilon_I}(i\Gamma^{23})\left(
F_{01}\Gamma^{01}+F_{23}\Gamma^{23}+F_{45}\Gamma^{45}+F_{67}\Gamma^{67}+F_{89}\Gamma^{89}
\right)\Gamma^3\Psi = 0,\quad (I=1,2).
\end{align}
The lefthand side is trivially satisfied for $\epsilon_1$ which satisfies
 $\overline{\epsilon_1}(i\Gamma^{M,M+1})=-\overline{\epsilon_1}$.
For $\epsilon_2$ this equation gives the condition
\begin{align}
 \Trp \overline{\epsilon_2}\Big(
F_{01}-i\left(
F_{23}+F_{45}+F_{67}+F_{89}
\right)
\Big)\Gamma^3\Psi = 0.
\end{align}
Then,
\begin{equation}
F_{01}=0,\qquad
F_{23}+F_{45}+F_{67}+F_{89}=0.
\end{equation}
The second equation \eqref{02SUSYCond2} of $(M,N)$$=$$(0,2)$, $(2,4)$, $(2,6)$ and $(2,8)$ are trivially satisfied for the case of $\epsilon_2$ in the same way and in the cases $(M,N)$$=$$(0,4)$,(0,6)$,(0,8)$,(4,6)$,(4,8)$ and $(6,8)$ this equation becomes trivial for $\epsilon_1$.

The condition for the supersymmetry generated by $\epsilon_I$ to be preserved is summarized as follows:
\paragraph{(i) Supersymmetry generated by $\epsilon_1$}
\begin{equation}\label{02SUSY2}
  \begin{cases}
   F_{0, M}+F_{1, M} = 0 & (M=2,3), \\
   F_{2, M}-F_{3, M+1} =  F_{2, M+1} + F_{3, M} = 0 & (M=4,6,8).
  \end{cases}
\end{equation}
\paragraph{(ii) Supersymmetry generated by $\epsilon_2$}
\begin{equation}\label{02SUSY1}
  \begin{cases}
   F_{0,1}=0, \:\: F_{23}+F_{45}+F_{67}+F_{89}=0\\
   F_{0, M}+F_{1, M} = 0 & (M=4,5,6,7,8,9), \\
   F_{M, N}-F_{M+1, N+1} =  F_{M, N+1} + F_{M+1, N} = 0 & ((M,N)=(4,6),(4,8),(6,8)).
  \end{cases}
\end{equation}

Let us define complex fields
\begin{equation}
Z_1 :=\Phi_1 + i\Phi_2,\:\:
Z_2 :=\Phi_3 + i\Phi_4,\:\:
Z_3 :=\Phi_5 + i\Phi_6.
\end{equation}
We define coordinates on the 2d CFT and the Riemann surface and redefine gauge field on them.
\begin{align}
\text{ On } (x^0, x^1) :\:\: & x^0 \pm x^1 =: x^\pm,\:\:
A_{x\pm} := \frac{1}{2}\;(A_0 \pm A_1),\\
\text{ On } (x^2, x^3) :\:\: & x^2 \pm i x^3 =: w^\pm,\;\;
A_{w\pm} := \frac{1}{2}\;(A_2 \mp i A_3).
\end{align}
Then, the following new derivatives can be defined:
 \begin{align}
 & \frac{1}{2}\:\Big(D_0 \pm D_1\Big)
  = \left(\frac{\partial}{\partial x^\pm} + [A_{x\pm}, *] \right)
  =: D_{x\pm},\\
 & \frac{1}{2}\:\Big(D_2 \mp i D_3\Big)
  = \left(\frac{\partial}{\partial w^\pm} + [A_{w\pm}, *]\right)
  =: D_{w\pm}.
 \end{align}
Using these notations the supersymmetry conditions \eqref{02SUSY2}, \eqref{02SUSY1} are respectively rewritten as follows.
%%%%%%%%%
\begin{enumerate}
\item{Supersymmetry generated by $\epsilon_1$:}
\begin{align}
\eqref{02SUSY2} &\Rightarrow
  \begin{cases}
F_{0M}+F_{1M}=0\:\: (M=2,3),\\
D_{w-}Z_A=0.\\
  \end{cases}
\end{align}
\item{Supersymmetry generated by $\epsilon_2$:}
\begin{align}
\eqref{02SUSY1} &\Rightarrow
  \begin{cases}
F_{01}=0,\:\: F_{23}=-\frac{i}{2}\sum_{i}[Z_i,\overline{Z}_i],\\
D_{x+} Z_i = 0,\\
[Z_i, Z_j] = 0.
   \end{cases}
\end{align}
\end{enumerate}
%%%%%%%%%
In the second case we find that this equation looks like a Hitchin system \cite{Hitchin01071987}.
For more details of these types of equations, see \cite{Gaiotto:2009hg}.

%%%%%(2,2) case
\subsubsection{$\mathcal{N}=(2,2)$ case}\label{Subsec:22case}
The supersymmetry parameters preserved in the bulk are $\epsilon_1,\dots,\epsilon_4$ in Table \ref{44EV}.
In this case we can use the same method to the previous $\mathcal{N}=(0,2)$ case.
The normal component of the current satisfies:
\begin{align}
& \Trp \:\overline\epsilon\left(
F_{01}\Gamma^{01}+F_{23}\Gamma^{23}+F_{45}\Gamma^{45}+F_{67}\Gamma^{67}+F_{89}\Gamma^{89}
\right)\Gamma^3\Psi = 0\label{22SUSYCond1},\\
& \Trp \:\overline\epsilon\left(
F_{M,N}\Gamma^{M,N}+F_{M,N+1}\Gamma^{M,N+1}+F_{M+1,N}\Gamma^{M+1,N}+F_{M+1, N+1}\Gamma^{M+1, N+1}
\right)\Gamma^3\Psi = 0.
\label{22SUSYCond2}\nonumber\\\end{align}
%%%%%%%%
%The gamma matrices are diagonalized as follows:
%\begin{align}
%  \left\{
%    \begin{array}{l}
%      \Gamma^{01} \epsilon_I = -\lambda^0_I \epsilon_I \\
%      i\Gamma^{M,M+1} \epsilon_I = -\lambda^M_I \epsilon_I
%    \end{array}
%  \right.\qquad\Leftrightarrow\qquad
%  \left\{
%    \begin{array}{l}
%      \overline{\epsilon_I}\Gamma^{01} = \lambda^0_I\overline{\epsilon_I} \\
%      \epsilon_I (i\Gamma^{M,M+1}) = \lambda^M_I \overline{\epsilon_I},
%    \end{array}
%  \right.
%\end{align}
%where eigenvalues $\lambda^0_I, \lambda^M_I$ take values $+1$ or $-1$ and are summarized in Table \ref{Table:22EV}.
%\begin{table}[h]
%\centering
%  \begin{tabular}{| l  c || c | c | c | c | c |}
%    \hline
%     & & $\lambda^{0}_{I}$ & $\lambda^{2}_{I}$ & $\lambda^{4}_{I}$ & $\lambda^{6}_{I}$ & $\lambda^{8}_{I}$ \\ \hline
%   \multirow{4}{*}{$I$} & 1 & $-$ & $+$ & $+$ & $+$ & $+$ \\
%   				& 2 & $-$ & $-$ & $-$ & $-$ & $-$  \\ \cline{3-7}
%      				& 3 & $+$ & $+$ & $+$ & $+$ & $-$  \\
%      				& 4 & $+$ & $-$ & $-$ & $-$ & $+$  \\ \hline
%  \end{tabular}
%\caption{Eigenvalues, $\lambda^M_I$ :  $\mathcal{N}=(2,2)$ case.}\label{Table:22EV}
%\end{table}
%

The first equation \eqref{22SUSYCond1} becomes trivial for $\epsilon_I$ having eigenvalue $\lambda^2_I=+1$ in the same way to $\mathcal{N}=(0,2)$ case and for $\epsilon_I$ having eigenvalue $\lambda^2_I=-1$ this equation becomes
\begin{align}
F_{01}=0,\:\: \sum_{i=1}^{4} \lambda^{2i}_I F^{2i,2i+1}=0.
\end{align}

The second equation \eqref{22SUSYCond2}  splits into two groups
\begin{align*}
(M,N) &= (0,2), (2,4), (2,6), (2,8), \\
(M,N) &= (0,4), (0,6), (0,8), (4,6), (4,8), (6,8).
\end{align*}
The former becomes trivial for $\lambda^2_I=-1$ and the latter becomes trivial for $\lambda^2_I=+1$.
The nontrivial conditions are for $\lambda^2_I=+1$
\begin{align}
& F_{02}-\lambda^0_I F_{12} = F_{03}-\lambda^0_I F_{13} = 0,\\
& F_{M,N}-\lambda^M_I\lambda^N_I F_{M+1,N+1}
= +\lambda^M_I F_{M+1,N}+\lambda^N_I F_{M,N+1}=0\nonumber\\
&\hspace{5cm} (M,N)=(2,4)(2,6)(2,8),
\end{align}
and for $\lambda^2_I=-1$
\begin{align}
& F_{0,M}-\lambda^0_IF_{1,M} = 0, \qquad M=4,5,6,7,8,9,\\
& F_{M,N}-\lambda^M_I\lambda^N_IF_{M+1,N+1}
= \lambda^M_IF_{M+1,N}+\lambda^N_IF_{M,N+1}=0\nonumber\\
&\hspace{5cm} (M,N)=(4,6)(4,8)(6,8).
\end{align}

Summarizing the above, the supersymmetries  generated by $\epsilon_I$ are respectively as follows:
\begin{enumerate}
\item{$\epsilon_I\:(\lambda^2_I=+1)$}
\begin{align}\label{Item:02breakcase}
\begin{cases}
F_{0,M}-\lambda^0_IF_{1,M}=0\qquad &(M=2,3)\\
F_{M,N}-\lambda^M_I\lambda^N_IF_{M+1,N+1}
= \lambda^M_IF_{M+1,N}+\lambda^N_IF_{M,N+1}=0
	& (M,N)=(2,4)(2,6)(2,8),
\end{cases}
\end{align}
\item{$\epsilon_I\:(\lambda^2_I=-1)$}
\begin{align}
\begin{cases}
F_{01}=0,\qquad \sum_{i=1}^{4}\lambda^{2i}_IF^{2i,2i+1}=0\\
F_{0,M}-\lambda^0_IF_{1,M}=0 &(M=4,5,6,7,8,9)\\
F_{M,N}-\lambda^M_I\lambda^N_IF_{M+1,N+1}
= \lambda^M_IF_{M+1,N}+\lambda^N_IF_{M,N+1}=0
& (M,N)=(4,6)(4,8)(6,8).
\end{cases}
\end{align}
\end{enumerate}

The case we studied before in the subsection \ref{Subsec:02case} corresponds to the case of $\lambda^0_I=-1$ (eqs.\eqref{Item:02breakcase}) in the current case.

%%%Subsec(4,4)case
\subsubsection{$\mathcal{N}=(4,4)$ case}\label{Subsec:44case}
The supersymmetry parameters preserved in the bulk are $\epsilon_1,\dots,\epsilon_8$ in Table \ref{44EV}.
The conditions for the bosonic fields are
\begin{align}
F_{0,1} = 0, \:\:
\sum_{i=1}^{4} \lambda_I^{2i} F_{2i,2i+1}=0,
\end{align}
where $I=2,4,6,8$ and
\begin{align}
F_{0M}-\lambda^0_IF_{1M}=0,
\end{align}
where $M=2,3$ for $I=2,4,6,8$ while $M=4,5,6,7,8,9$ for $I=1,3,5,7$, and
\begin{align}
F_{M,N}-\lambda_I^M\lambda_I^NF_{M+1,N+1}
	= \lambda_I^N F_{M,N}-\lambda_I^M F_{M+1,N+1} = 0,
\end{align}
where $(M,N)=(2,4),(2,6)$ and $(2,8)$ for $I=$ even,
while $(M,N)=(4,6),(4,8)$ and $(6,8)$ for $I=$ odd.
The case of $\mathcal{N}=(2,2)$ with the boundary is an interesting case and the central charge is obtained only from the calculation of the 't Hooft anomaly, as shown in subsection \ref{Subsec:central-charge}.

\section*{Acknowledgement} 
We are pleased to thank Dongmin Gang,  Kentaro Hori, Masahito Yamazaki, and Yutaka Yoshida
for useful discussion.
K.N. would like to thank the organizers of ``the 7th Taiwan String Workshop'' at National Taiwan University for giving him an opportunity to talk about this work.
K.N. was supported in part by JSPS Research Fellowship for Young Scientists and JSPS KAKENHI Grant Number 13J02068.
This work was supported in part by World Premier International Research Center Initiative (WPI), MEXT, Japan.

\providecommand{\href}[2]{#2}\begingroup\raggedright\endgroup
\end{document}